\documentstyle[12pt,axodraw,equation,epsfig]{article}

\newcommand{\imag}{\Im {\rm m}}
\newcommand{\real}{\Re {\rm e}}

\begin{document}

\renewcommand{\thefootnote}{\fnsymbol{footnote}}

\mbox{ }\\[-1cm]
\mbox{ }\hfill SNU--TP 00--038\\
\mbox{ }\hfill \today\\

\vskip 1cm

\begin{center}
{\Large\bf Neutralino--Nucleus Elastic Cross Section in the Minimal 
           Supersymmetric Standard Model with Explicit CP Violation}\\[1cm]   
S.Y. Choi $^1$,\, Seong Chan Park $^2$,\, J.H. Jang $^3$ and H.S. Song $^2$
\end{center}

\bigskip 

\begin{center}
$^1${\it Department of Physics, Chonbuk National University, 
          Chonju 561--756, Korea}\\[1mm]
$^2${\it Department of Physics and CTP, 
         Seoul National University, Seoul 151-742, Korea}\\[1mm]
$^3${\it Institute of Phontonics, Electronics and Information 
         Technology\\
	 Chonbuk National University, Chonju 561--756, Korea}
\end{center}

\vskip 2.5cm

\begin{abstract}
We study the elastic scattering of the lightest neutralino with a  
nucleus in the framework of the minimal supersymmetric standard model 
(MSSM) with explicit flavor preserving CP violation, including the
one--loop CP--violating neutral Higgs--boson mixing effects induced 
dominantly by the CP phases in the top and bottom (s)quark sectors.
We construct the most general form of the effective Lagrangian for 
the neutralino--nucleus scattering in the limit of vanishing momentum 
transfers and then we perform a comprehensive analysis of the effects 
of the complex CP phases on the mass spectra of the lightest neutralino, 
neutral Higgs bosons and top squarks, and on the the spin--dependent
and spin--independent neutralino--nucleus scattering cross section 
for three neucleus targets F, Si and Ge. The CP phases can reduce or
enhance the neutralino--nucleus cross sections significantly,
depending on the values of the real parameters in the MSSM.
\vskip 0.5cm

\noindent
PACS number(s): 11.30.Pb, 11.30.Er
\end{abstract}

\newpage

\section{Introduction}
\label{sec:section 1}

A lot of cosmological and astrophysical observations have revealed that 
the dominant fraction of the matter in the universe and in the galactic halo 
is not luminous or dark \cite{Kolb:1990vq}.  Furthermore, big--bang 
nucleosynthesis and galactic structure formation require some non--baryonic 
and non--relativistic matter for which various particle physics theories,
in particular, supersymmetric theories provide good candidates 
\cite{Nilles:1984ge,Jungman:1996df}. The lightest supersymmetric particle 
(LSP), which is guaranteed by R--parity conservation, is one of the most 
well--established and studied weakly interacting massive particle (WIMP) 
candidates of dark matter(DM). The null experimental results in searching 
for the anomalously heavy isotopes and the considerations in the most models 
suggest that the LSP in the minimal supersymmetric standard model (MSSM)
might be the lightest neutralino, a linear combination of the four neutral 
superpartners of the U(1)$_Y$ and SU(2)$_L$ gauge bosons and of the two 
neutral Higgs bosons;
\begin{eqnarray}
\tilde{\chi}_1^0\, =\, N_{11}\,\widetilde{B}+N_{12}\, \widetilde{W}^3 
                   +   N_{13}\,\widetilde{H}^0_1+N_{14}\, \widetilde{H}^0_2,
\end{eqnarray} 
where $N$ is the unitary matrix diagonalizing the $4\times4$ neutralino mass 
matrix \cite{Ellis:1984ew}.\\

The lightest neutralino as a dominant cold dark matter (CDM) has very 
weak, but finite coupling strengths to quarks so that there might be a chance 
to detect the CDM neutralino even in laboratories directly or 
indirectly \cite{Primack:1988zm}. If the couplings are too small, the 
neutralinos would not have annihilated in the early universe and it would be 
too much abundant today. Certainly, the capabilities of both direct and 
indirect search experiments for the lightest neutralino rely crucially on
the size of the elastic neutralino--matter scattering cross section, in
particular the neutralino--nucleus cross section because the detection rates 
for either direct or indirect searches are proportional to the cross sections. 
Since the first rough estimates were made by Goodman and 
Witten \cite{Goodman:1985dc}, the estimates of the neutralino--nucleus 
scattering cross section have been continuously and significantly 
improved \cite{Griest:1988ma,FGC}. Furthermore, there are 
many on--going and planned experiments with the goal of direct or indirect 
detections of the CDM neutralinos among which some experiments have already 
reported interesting but controversial results \cite{Bernabei:1999td}. 
Therefore, it is very important and timely to evaluate the cross section 
for neutralino--nucleus elastic scattering by including all the possibly
dominant factors in determining the elastic scattering cross section
as well as the neutralino--neutralino annihilation cross sections in
the supersymmetric theories.\\

In the present work we re--investigate the elastic scattering of the
neutralino--nucleus scattering in the framework of the minimal supersymmetric 
standard model (MSSM) with R--parity focusing on all the dominant effects 
of flavor--preserving CP--violating complex phases on the cross section
\cite{FGC}.
The effects of the phase $\Phi _\mu$ of the higgsino mass parameter $\mu$ 
on the neutralino--nucleus scattering as well as the neutralino relic density
have been already studied by several works \cite{Falk}. 
However, there can exist additional effects through two different types of 
CP--violating phases\footnote{Without any loss of generality, the 
SU(2)$_L$ gaugino mass $M_2$ can be taken to be real}; one is the phase 
$\Phi_1$ of the U(1)$_Y$ gaugino mass $M_1$ and the other are the phases 
of the trilinear terms $A_f$ in the sfermion mass matrices. The latter phases
can cause the CP--violating mixing among three neutral Higgs bosons at 
one--loop level so that the couplings of the Higgs bosons to the 
(s)particles are significantly modified \cite{CEPW,Choi:2000wz,Demir}. 
We note that the phases of the trilinear parameters for the first and second 
generation sfermions are strongly constrained by the stringent experimental 
bounds on the electron and neutron electric dipole moments (EDMs) unless the 
sfermions are too heavy. On the other hand, since the third generation 
sfermion sectors give only two--loop suppressed effects on the electron and 
neutron EDMs when any generational mixing among sfermions is neglected, 
the CP violating phases involving the third generation sfermions can be 
very large \cite{CEPW}. In addition, the most dominant contributions 
to the CP--violating neutral Higgs boson mixing are from the third generation 
sfermion sectors because of their large Yukawa couplings \cite{CEPW,Choi:2000wz}. 
So, including the CP--violating neutral Higgs boson mixing due to the phases 
of the third generation trilinear parameters and the phase $\Phi_1$ as well 
as the phase $\Phi_\mu$ we will provide a comprehensive analysis of the 
dependence of the neutralino--nucleus elastic scattering cross section 
on those phases in the present work.\\

The typical size of the momentum transfers exchanged in the 
neutralino--nucleus elastic scattering is of the order of 10 KeV, much 
smaller than the neutralino mass of the order of 100 GeV. For such tiny 
momentum transfers, the interactions of the lightest neutralino with 
spin--1/2 quarks can be described by two effective four--Fermi 
current$\times$ current terms; the spin--dependent 
axial--vector$\times$axial--vector four--Fermi Lagrangian  
and the spin--independent scalar$\times$scalar four--Fermi Lagrangian.
%
%
%
We note that the spin--independent term becomes more significant for the 
scattering of the neutralino off nuclei with a large mass number because
each spin--independent neutralino--nucleon scattering contributes coherently 
to the overall spin--independent cross section. 
As a result, the spin--independent part dominates the spin--dependent part 
for large target nuclei.\\

In the CP--noninvariant case, the scalar---pseudoscalar mixing among neutral
Higgs bosons in the MSSM modifies the scalar couplings of each neutral Higgs 
bosons to fermions and neutralinos, leading to the significant changes of
the spin--independent neutralino--nucleus cross section. It is, therefore,
worthwhile to make a systematic investigation of the effects of all
the relevant CP--violating phases on the neutralino--nucleus scattering 
cross section in the framework of the MSSM. For a systematic quantitative 
investigation, our analysis in the present work is based on a specific 
framework with the following assumptions:
\begin{itemize}
\item The first and second generation sfermions are assumed to be very heavy 
      so that they are decoupled from the theory. In this case, there are 
      no constraints on the CP--violating phases from the neutron and 
      EDMs. On the other hand, the 
      annihilation of the neutralinos into tau pairs through the exchange 
      of relatively light scalar tau leptons guarantees that the 
      cosmological constraints on the DM densities be satisfied.
\item The explicit CP violation in the Higgs sector induced through the 
      CP--violating radiative corrections from the top and bottom squark 
      sectors is included.
\item Simultaneously, the effects of the induced CP--violating phase between 
      Higgs doublets on the chargino and neutralino systems are explicitly 
      included.
\item It is necessary to avoid the possible constraints from the so--called 
      Barr--Zee--type diagrams to the electron and neutron EDMs as well as 
      from the Higgs search experiments at LEP. We take two 
      values 3 and 10 for $\tan\beta$, the ratio of the vacuum expectation
      values of two neutral Higgs fields. 
\end{itemize}
\mbox{ }\\[-4mm]

The rest of the paper is organized as follows. In Sec.~\ref{sec:section 2}, 
we present a brief review of the CP--violating mixing among scalar and
pseudo--scalar Higgs bosons in the MSSM Higgs sector on the basis of the
recent work by Choi, Drees and Lee \cite{Choi:2000wz}. In addition, we
discuss the neutralino mixing including the CP--violating phase induced from
the MSSM Higgs sector.  In Sec.~\ref{sec:section 3}, we derive the complete
analytic expressions for the spin--dependent and spin--independent effective 
four--Fermi Lagrangian relevant to our calculations explicitly, taking into 
account all the Higgs--boson, scalar--quark and $Z$--boson exchange diagrams 
in the limit of vanishing momentum transfers. Section~\ref{sec:section 4}
is devoted to a detailed numerical analysis for investigating 
the dependence of the neutralino--nucleus elastic scattering
cross section for the three target nuclei F, Si, and Ge on the CP violating 
phases as well as other relevant real SUSY parameters 
such as $\tan \beta$ by taking a few typical parameter sets which effectively
cover the whole parameter space. Finally, we summarize our 
findings and conclude in Sec.~\ref{sec:section 5}.

\section{CP--violating mixing in the MSSM}
\label{sec:section 2}

\subsection{Neutral Higgs--boson mixing}

In this section, we give a brief review of the calculation 
\cite{Choi:2000wz} of the Higgs-boson mass matrix based on the full 
one-loop effective potential, valid for all values of the relevant 
third--generation soft--breaking parameters.\\

The MSSM contains two Higgs doublets $H_1, \ H_2$, with hypercharges
$Y(H_1) = -Y(H_2) = -1/2$. Here we are only interested in the neutral
components, which we write as
\begin{eqnarray}
\label{e1}
H_1^0=\frac{1}{\sqrt{2}}\left(\phi_1 + i\, a_1\right), \qquad
H_2^0=\frac{{\rm e}^{i \xi}}{\sqrt{2}}\left(\phi_2 + i\, a_2\right),
\end{eqnarray}
where $\phi_{1,2}$ and $a_{1,2}$ are real scalar fields. The constant phase
$\xi$ can be set to zero at tree level, but will in general become
non--zero once loop corrections are included. 
The mass matrix of the neutral Higgs bosons can be computed from the
effective potential \cite{101}
\begin{eqnarray}
\label{e2}
V_{\rm Higgs} \hskip -0.6cm
   &= \, \frac{1}{2}m_1^2\left(\phi_1^2+a_1^2\right)
     +\frac{1}{2}m_2^2\left(\phi_2^2+a_2^2\right)
     -\left|m_{12}^2\right|\left(\phi_1\phi_2-a_1 a_2\right) 
           \, \cos (\xi + \theta_{12}) \nonumber \\ 
   & +\left|m^2_{12}\right|
      \left(\phi_1 a_2 +\phi_2 a_1\right)\,\sin(\xi+\theta_{12})
     +\frac{\hat{g}^2}{8} {\cal D}^2 
     +\frac{1}{64\,\pi^2}\,\, {\rm Str}\! \left[
           {\cal M}^4 \left(\log\frac{{\cal M}^2}{Q^2} 
	                  - \frac{3}{2}\right)\right],
\end{eqnarray}
where we have allowed the soft breaking parameter $m^2_{12} = \left|
\, m^2_{12} \right|\, {\rm e}^{i \theta_{12}}$ to be complex, and we have
introduced the quantities
\begin{eqnarray}
\label{e5}
{\cal D} = \phi_2^2 + a_2^2 - \phi_1^2 - a_1^2\, , \qquad \hat{g}^2 
         = \frac{g^2 + g'^2}{4},
\end{eqnarray}
where the symbols $g$ and $g'$ stand for the SU(2)$_L$ and U(1)$_Y$
gauge couplings, respectively. $Q$ in Eq.~(\ref{e2}) is the
renormalization scale; the parameters of the tree--level potential, in
particular the parameters $m_1^2, \ m_2^2 $ and $m_{12}^2$, are
running masses, taken at scale $Q$. The potential (\ref{e2}) is
then independent of $Q$, up to two--loop corrections.\\

The matrix ${\cal M}$ in Eq.~(\ref{e2}) is the field--dependent mass matrix 
of all modes that couple to the Higgs bosons. The by far dominant 
contributions come from the third generation quarks and squarks. 
The (real) masses of the former are given by
\begin{eqnarray}
\label{e3}
m_b^2 = \frac{1}{2}\, |h_b|^2 \left(\phi_1^2 + a_1^2\,\right)\,, \qquad
m_t^2 = \frac{1}{2}\, |h_t|^2 \left(\phi_2^2 + a_2^2\,\right), 
\end{eqnarray}
where $h_b$ and $h_t$ are the bottom and top Yukawa couplings. The
corresponding squark mass matrices can be written as
\begin{eqnarray}
\label{sqmass}
&& {\cal M}_{\tilde t}^2 = \left(\begin{array}{cc} 
  m^2_{\widetilde Q}+m_t^2-\frac{1}{8}\left(g^2-\frac{g'^2}{3}\right){\cal D}
 &- h_t^* \left[ A_t^* \left(H_2^0 \right)^* + \mu H_1^0\right] \\[2mm]
  - h_t \left[ A_t H^0_2 + \mu^* \left( H_1^0 \right)^* \right] 
 &  m^2_{\widetilde U} + m_t^2 - \frac{g'^2}{6} {\cal D}
                                 \end{array} \right),  \nonumber\\[3mm]
&& {\cal M}_{\tilde b}^2 = \left( \begin{array}{cc} 
  m^2_{\widetilde Q}+m_b^2+\frac{1}{8}\left(g^2+\frac{g'^2}{3}\right){\cal D}
 &- h_b^* \left[ A_b^*  \left( H_1^0 \right)^* + \mu H_2^0 \right] \\[2mm]
  - h_b \left[ A_b H_1^0 + \mu^* \left( H_2^0 \right)^* \right] 
 &  m^2_{\widetilde D} + m_b^2 + \frac{g'^2}{12} {\cal D}
\end{array} \right).  
\label{e4}
\end{eqnarray}
Here, $H_1^0$ and $H_2^0$ are given by Eq.~(\ref{e1}) while $m_t^2$
and $m_b^2$ are as in Eq.~(\ref{e3}) and ${\cal D}$ has been defined
in Eq.~(\ref{e5}). In Eq.~(\ref{e4}) $m^2_{\widetilde Q}, \ 
m^2_{\widetilde U}$ and
$m^2_{\widetilde D}$ are real soft breaking parameters, $A_b$ and $A_t$ are
complex soft breaking parameters, and $\mu$ is the complex
supersymmetric Higgs(ino) mass parameter.  \\

The mass matrix of the neutral Higgs bosons can now be computed from
the matrix of second derivatives of the potential (\ref{e2}), where
(after taking the derivatives) $m_1^2, \ m_2^2$ and $m_{12}^2 \sin(
\xi + \theta_{12})$ are determined by the stationarity conditions.
The massless state $G^0 = a_1 \cos \beta - a_2 \sin \beta$ is
the would--be Goldstone mode ``eaten'' by the longitudinal $Z$ boson. 
We are thus left with a
squared mass matrix ${\cal M}_H^2$ for the three states $a = a_1 \sin
\beta + a_2 \cos \beta, \ \phi_1$ and $\phi_2$. This matrix is real
and symmetric, i.e. it has 6 independent entries. 
The diagonal entry for $a$ reads:
\begin{eqnarray}
\label{e10}
\left. {\cal M}^2_{H} \right|_{aa} = m_A^2 + \frac {3} {8 \pi^2}
\left\{ \frac { |h_t|^2 m_t^2 } { \sin^2 \beta} g(m^2_{\tilde{t}_1},
 m^2_{\tilde{t}_2}) \Delta_{\tilde t}^2 + 
\frac {|h_b|^2 m_b^2 } { \cos^2 \beta} g(m^2_{\tilde{b}_1},
 m^2_{\tilde{b}_2}) \Delta_{\tilde b}^2 \right\},
\end{eqnarray}
and the CP--violating entries of the mass matrix,
which mix $a$ with $\phi_1$ and $\phi_2$ read:
\begin{eqnarray}
\label{e13} 
\left. {\cal M}^2_H \right|_{a \phi_1} 
  &=& \frac {3} {16 \pi^2} \left\{
  \frac { m_t^2 \Delta_{\tilde t} } {\sin \beta} \left[ g(m^2_{\tilde{t}_1}, 
        m^2_{\tilde{t}_2})
  \left( X_t \cot\!\beta - 2 \left|h_t\right|^2 R_t \right) 
          - \hat{g}^2\cot\!\beta\log
	    \frac{m^2_{\tilde{t}_2}}{m^2_{\tilde{t}_1}} \right] \right.
	    \nonumber\\  
  && \left. \hskip 1cm
  +\frac {m_b^2 \Delta_{\tilde b}} {\cos \beta} 
   \left[ -g(m^2_{\tilde{b}_1},m^2_{\tilde{b}_2})
  \left( X_b + 2 \left| h_b\right|^2 R_b' \right) 
  + \left( \hat{g}^2 - 2 \left| h_b\right|^2 \right) \log
     \frac {m^2_{\tilde{b}_2}}{m^2_{\tilde{b}_1}}\right] \right\}, \nonumber\\
\left. {\cal M}^2_H \right|_{a \phi_2} 
   &=& \frac {3} {16 \pi^2} \left\{
    \frac {m_t^2 \Delta_{\tilde t}} {\sin \beta} 
     \left[ -g(m^2_{\tilde{t}_1}, m^2_{\tilde{t}_2})
   \left( X_t + 2 \left| h_t\right|^2 R_t' \right) 
   + \left(\hat{g}^2-2\left| h_t\right|^2\right)
     \log\frac{m^2_{\tilde{t}_2}}{m^2_{\tilde{t}_1}} \right]\right. 
     \nonumber\\ 
  && \left. \hskip 1cm
  +\frac { m_b^2 \Delta_{\tilde b} } {\cos \beta} 
   \left[ g(m^2_{\tilde{b}_1}, m^2_{\tilde{b}_2})
  \left( X_b \tan\!\beta - 2 \left| h_b\right|^2 R_b \right) 
   - \hat{g}^2\tan\!\beta\log 
     \frac{m^2_{\tilde{b}_2}}{m^2_{\tilde{b}_1}}\right]\right\}. 
\end{eqnarray}
where $\Delta_{\tilde t}$ and $\Delta_{\tilde b}$, which 
describe the amount of CP violation in the squark mass, read
\begin{eqnarray}
\label{e9}
\Delta_{\tilde t}=\frac{ \imag(A_t \mu {\rm e}^{i \xi}) }
                  {m^2_{\tilde{t}_2} - m^2_{\tilde{t}_1}} ; \qquad
\Delta_{\tilde b}=\frac{ \imag(A_b \mu {\rm e}^{i \xi}) }
                  {m^2_{\tilde{b}_2} - m^2_{\tilde{b}_1}} ,
\end{eqnarray}
and $g(x,y) = 2 - [(x+y)/(x-y)]\, \log (x/y)$.
The definition of the mass squared $m^2_A$ and the dimensionless 
quantities $X_{t,b}$, $R_{t,b}$ and $R^\prime_{t,b}$ as well as the other 
CP--preserving entries of 
the mass matrix squared ${\cal M}^2_{H}$  can be found 
in Ref.~\cite{Choi:2000wz}. As noted earlier, the size of these 
CP--violating entries is controlled by
$\Delta_{\tilde t}$ and $\Delta_{\tilde b}$. \\

The real and symmetric $3\times 3$ matrix ${\cal M}^2_H$ can be
diagonalized by an $3\times 3$ rotation $O$;
\begin{eqnarray}
\left(\begin{array}{c}
      a \\
      \phi_1 \\
      \phi_2
      \end{array}\right) = O\,
\left(\begin{array}{c}
       H_1 \\
       H_2 \\
       H_3
      \end{array}\right) 
\end{eqnarray}
with the increasing order of the three mass eigenvalues, 
$m^2_{H_1} \leq m^2_{H_2}\leq m^2_{H_3}$, taken as a convention. 
Note that the loop--corrected neutral Higgs--boson sector is determined 
by fixing the values of various parameters; $m_A$, $\mu$, $A_t$, $A_b$, 
a renormalization scale $Q$, $\tan\beta$, and the soft--breaking third 
generation sfermion masses, $m_{\tilde Q}$, $m_{\tilde U}$, 
and $m_{\tilde D}$. The radiatively induced phase $\xi$ is no more 
an independent parameter and it can be absorbed into the 
definition of the $\mu$ parameter so that the physically meaningful
CP phases in the Higgs sector are the phases of the re--phasing 
invariant combinations $A_t \mu {\rm e}^{i\xi}$ and 
$A_b \mu {\rm e}^{i\xi}$. This neutral Higgs--boson mixing changes the 
couplings of the Higgs fields to fermions, gauge bosons, and 
Higgs fields themselves so that the effects of CP violation in 
the Higgs sector can be probed through various 
processes \cite{EXCP_FC}.\\

\subsection{Top and bottom squark mixing}

On the other hand, the $2\times 2$ Hermitian top and bottom squark mass 
matrices squared ${\cal M}_{{\tilde t}}^2$ and ${\cal M}_{{\tilde b}}^2$ 
can be obtained after plugging the vacuum expectation values for the Higgs 
fields:
\begin{eqnarray}
&& {\cal M}^2_{\tilde{t}}
    =\left(\begin{array}{cc}
      m^2_{\tilde{Q}}+m^2_t+m^2_Z\,\cos 2\beta\,
      \left(\frac{1}{2}-\frac{2}{3}\, s^2_W\right) & 
      -m_t\left(A^*_t+\mu\, {\rm e}^{i\xi}\, \cot\beta\right)\\[2mm]
      -m_t\left(A_t+\mu^*\, {\rm e}^{-i\xi}\, \cot\beta\right) &
      m^2_{\tilde{U}}+m^2_t+\frac{2}{3} m^2_Z\,\cos 2\beta\, s^2_W 
           \end{array}\right),\nonumber\\[3mm]
&& {\cal M}^2_{\tilde{b}}
    =\left(\begin{array}{cc}
      m^2_{\tilde{Q}}+m^2_b-m^2_Z\,\cos 2\beta\,
      \left(\frac{1}{2}+\frac{1}{3}\, s^2_W\right) & 
      -m_b\left(A^*_b+\mu\, {\rm e}^{i\xi}\, \tan\beta\right)\\[2mm]
      -m_b\left(A_b+\mu^*\, {\rm e}^{-i\xi}\, \tan\beta\right) &
      m^2_{\tilde{D}}+m^2_b-\frac{1}{3} m^2_Z\,\cos 2\beta\, s^2_W 
           \end{array}\right)
\end{eqnarray}
These Hermitian mass matrices can be diagonalized by the unitary matrices, 
$U_{{\tilde t}}$ and $U_{{\tilde b}}$,
\begin{eqnarray}
\left(\begin{array}{c}
      \tilde{t}_L \\
      \tilde{t}_R
      \end{array}\right)= U_{\tilde{t}}\,
\left(\begin{array}{c}
      \tilde{t}_1 \\
      \tilde{t}_2
      \end{array}\right)\,, \qquad
\left(\begin{array}{c}
      \tilde{b}_L \\
      \tilde{b}_R
      \end{array}\right)= U_{\tilde{t}}\,
\left(\begin{array}{c}
      \tilde{b}_1 \\
      \tilde{b}_2
      \end{array}\right)\,, 
\end{eqnarray}
respectively, where $\tilde{t}_{1,2}$ and $\tilde{b}_{1,2}$ are the mass 
eigenstates with their masses, $m_{\tilde{t}_{1,2}}$ and 
$m_{\tilde{b}_{1,2}}$,\,  respectively. We note that the off--diagonal terms
of each mass matrix squared are proportional to the corresponding fermion
mass so that their contributions can be significant only for the third
generation sfermions -- top and bottom squarks. \\

\subsection{Neutralino mixing}

In general, the induced phase $\xi$ in Eq.~(\ref{e1}) remains as a 
non--trivial physical phase and leads to a modification in the neutralino 
mass matrix; analytically, the induced phase plays a role of rotating the
vacuum expectation value $v_2$ into $v_2\, e^{i \xi}$ so that it modifies
the neutralino mass matrix describing the gauginos and higgsinos 
through electroweak symmetry breaking. In the weak--interaction basis 
$(\tilde{B},\, \tilde{W}^3,\, \tilde{H}_1^0,\, \tilde{H}_2^0 )$\, the $4\times4$
symmetric, but complex neutralino mass matrix reads 
\begin{eqnarray}
{\cal M}_N=\left(\begin{array}{cccc}
   |M_1|\, {\rm e}^{i\Phi_1}  &              0                     
 &   -m_Z\, c_\beta\, s_W     & m_Z\, s_\beta\, s_W\, {\rm e}^{-i\xi} \\[1mm]
           0                  &             M_2         
 &    m_Z\, c_\beta\, c_W     & -m_Z\, s_\beta\, c_W\, {\rm e}^{-i\xi} \\[1mm]
     -m_Z\, c_\beta\, s_W     &  m_Z\, c_\beta\, c_W
 &         0                  & -|\mu|\, {\rm e}^{i\Phi_\mu}  \\[1mm] 
     m_Z\, s_\beta\, s_W\, {\rm e}^{-i\xi} 
 &  -m_Z\, s_\beta\, c_W\, {\rm e}^{-i\xi}
 &  -|\mu|\, {\rm e}^{i\Phi_\mu} &              0
                \end{array}\right),
\label{eq:neutralino mass matrix}
\end{eqnarray}
where the phases of the U(1) gaugino mass $M_1$ and the higgsino mass
parameter $\mu$, $\{\Phi_1,\, \Phi_\mu\}$ are explicitly given.
The neutralino mass matrix is diagonalized in a symmetric way
through a unitary matrix $N$;
\begin{eqnarray}
\left(\begin{array}{c}
      \tilde{\chi}^0_1 \\[1mm]
      \tilde{\chi}^0_2 \\[1mm]
      \tilde{\chi}^0_3 \\[1mm]
      \tilde{\chi}^0_4 
      \end{array}\right) = N \, 
\left(\begin{array}{l}
      \tilde{B}     \\[1mm]
      \tilde{W}^3   \\[1mm]
      \tilde{H}^0_1 \\[1mm]
      \tilde{H}^0_2 
      \end{array}\right)\,,
\end{eqnarray}
with the increasing order of the mass eigenvalues, 
$m_{\tilde{\chi}^0_1}\leq m_{\tilde{\chi}^0_2}\leq m_{\tilde{\chi}^0_3}
\leq m_{\tilde{\chi}^0_4}$, as a convention. One can find that after 
an appropriate field redefinition there are two rephrasing--invariant 
phases $\Phi_1$ and $\Phi_\mu +\xi$ in the neutralino sector, 
where $\Phi_1$ is the phase of the U(1) gaugino mass $M_1$ and $\Phi_\mu$ 
the phase of the the higgsino mass parameter $\mu$.\\ 

\vspace*{0.5cm} 
\begin{center}
\begin{picture}(300,100)(0,0)


\Text(-25,100)[]{$\tilde{\chi}^0_1$}
\Text(50,100)[]{$\tilde{\chi}^0_1$}
\Text(-25,0)[]{$q$}
\Text(50,0)[]{$q$}
\Text(30,50)[]{$Z$}

\Line(-30,90)(12.5,70)
\Line(50,90)(12.5,70)
\Photon(12.5,70)(12.5,30){3}{8}
\ArrowLine(-25,10)(12.5,30)
\ArrowLine(12.5,30)(50,10)


\Text(110,100)[]{$\tilde{\chi}^0_1$}
\Text(185,100)[]{$\tilde{\chi}^0_1$}
\Text(110,0)[]{$q$}
\Text(185,0)[]{$q$}
\Text(165,50)[]{$H_k$}

\Line(105,90)(147.5,70)
\Line(185,90)(147.5,70)
\DashLine(147.5,70)(147.5,30){3}
\ArrowLine(110,10)(147.5,30)
\ArrowLine(147.5,30)(185,10)


\Text(245,85)[]{$\tilde{\chi}^0_1$}
\Line(250,75)(275,50)
\ArrowLine(250,25)(275,50)
\Text(245,15)[]{$q$}

\DashArrowLine(275,50)(315,50){3}
\Text(295,37)[]{$\tilde{q}_{1,2}$}

\Line(315,50)(340,75)
\Text(345,85)[]{$\tilde{\chi}^0_1$}
\ArrowLine(315,50)(340,25)
\Text(345,15)[]{$q$}

\end{picture}\\
\end{center}
\smallskip
\noindent
{\bf Figure~1}: {\it The six mechanisms contributing to the neutralino--quark 
                     elastic scattering process $\tilde{\chi}^0_1\, q 
		     \rightarrow\tilde{\chi}^0_1\, q$; the spin--1 $Z$ 
		     exchange, the three spin--0 neutral--Higgs--boson 
		     exchanges and the two squark $\tilde{q}_{1,2}$ 
		     exchanges. Here, the index $k$ denotes 1,2 or 3.}
\bigskip 
\bigskip 

\section{Neutralino--nucleus elastic cross section}
\label{sec:section 3}

\subsection{Feynman rules}

In this section, we present all the Feynman rules in terms of mass eigenstates
that are necessary for the neutralino--quark elastic scattering process (see
Fig.~1). \\

The interactions of the neutral gauge bosons $Z$ to quarks are
described by the same Lagrangian as given in the SM:
\begin{eqnarray}
{\cal L}_{Z q q}=\frac{e}{s_Wc_W}\,\bar{q}\, \gamma_\mu 
       \left[(Q_q\, s^2_W-T_3)\,\, P_L +Q_q\, s^2_W\, P_R \right] q\,  Z^\mu ,
\end{eqnarray}
where $Q_q$ and $T_3$ are the electric charge and the isospin of the quark,
respectively. The interactions of the neutral Higgs bosons with quarks 
are described by the Lagrangian
\begin{eqnarray}
\label{eq:higgs-quark}
{\cal L}_{H_k qq} =-\frac{h_d}{\sqrt{2}}\,\,\bar{d}
        \left[\,O_{2k} + i\,O_{1k}\,s_{\beta}\gamma_5 \right] d\, H_k  
                   -\frac{h_u}{\sqrt{2}}\,\,\bar{u}
        \left[\,O_{3k} + i\,O_{1k}\,c_{\beta}\gamma_5 \right] u\, H_k , 
\end{eqnarray}
where the Yukawa couplings of down-- and up--type quarks are given by
\begin{eqnarray}
h_d = \frac{e\, m_d}{\sqrt{2}\, s_W m_W c_{\beta}}, \qquad
h_u = \frac{e\, m_u}{\sqrt{2}\, s_W m_W s_{\beta}}.
\end{eqnarray}
On the other hand, the SU(2) and U(1) gauginos, $\tilde{W}^3$ and 
$\tilde{B}$, have zero isospin and electric charges so that they do not
couple to the $Z$ boson. 
However, the Higgsinos $\tilde{H}^0_1$ and $\tilde{H}^0_2$ have non--trivial
isospins so that the couplings of the $Z$ boson to the lightest neutralinos,
which are Majorana fermions, are described by the Lagrangian 
\begin{eqnarray}
{\cal L}_{Z\chi\chi} = \frac{e}{4 s_W c_W}
             \, \left[\, |N_{13} |^{\,2} - |N_{14} |^{\,2} \right]\,
   \left(\bar{\tilde{\chi}}^0_1\gamma^{\mu}\gamma_5\tilde{\chi}^0_1\right)
         	Z_{\mu}\,.
\end{eqnarray}
where the Majorana nature of the neutralino is reflected in the fact that
only the axial vector coupling but not the vector coupling is allowed. \\ 
   
On the other hand, the interactions of the lightest neutralino with a
pair of quark and squark come from both the gauge and Yukawa
interactions. We consider the general flavor--diagonal squark mixing 
as well as the neutralino mixing and then we obtain the following
Lagrangian   
\begin{eqnarray}
{\cal L}_{\chi\tilde{q}q} = -\frac{e}{\sqrt{2}s_W}
              \left\{\,\tilde{q}^*_1\bar{\tilde{\chi}^0}
                     \left[\,B^{1L}_{q} P_L + B^{1R}_{q}\,P_R\, \right]q
              + (1\rightarrow2)\right\}+{\rm H.c.}\,,
\end{eqnarray}
where the coefficients $B^{iL,R}_q$ ($i=1,\, 2$) in the mass eigenstate 
basis expressed in terms of the squark mixing angle $\theta_q$ and phase 
$\phi_q$ read
\begin{eqnarray}
&& B^{1L}_{q} = \cos\theta_{q}\, A^{LL}_{q} 
             + e^{-i\phi_{q}}\sin\theta_{q}\, A^{RL}_{q},\quad \,
   B^{1R}_{q} = \cos\theta_{q}\, A^{LR}_{q}
             + e^{-i\phi_{q}}\, \sin\theta_{q}\, A^{RR}_{q}, \nonumber \\[2mm]
&& B^{2L}_{q} = \cos\theta_{q}\, A^{RL}_{q}
             - e^{i\phi_{q}}\, \sin\theta_{q}\, A^{LL}_{q},\quad \hskip 0.3cm
   B^{2R}_{q} = \cos\theta_{q}A^{RR}_{q}
             - e^{i\phi_{q}}\, \sin\theta_{q}\, A^{LR}_{q} \,,
\end{eqnarray}
with the coefficients
\begin{eqnarray}
A^{LL}_{u}&=& N^*_{12} + \frac{1}{3}\, t_W\,N^*_{11},~~~~~~ 
A^{LR}_{u} = \frac{\sqrt{2}\, s_W h_u N_{14}}{e},    \nonumber \\[1mm]
A^{RL}_{u}&=& \frac{\sqrt{2}\, s_W h_u N^*_{14}}{e},~~~~~~~~
A^{RR}_{u} = \frac{4}{3}\, t_W\,N_{11} \,,
\end{eqnarray}
for the up--type quarks $u$, $c$ and $t$
and 
\begin{eqnarray}
A^{LL}_{d}&=& -N^*_{12} + \frac{1}{3}\, t_W\,N^*_{11},~~~~ 
A^{LR}_{d} = \frac{\sqrt{2}\, s_W h_d N_{13 }}{e}, \nonumber \\[1mm]
A^{RL}_{d}&=& \frac{\sqrt{2}\,s_W  h_d N^*_{13}}{e},~~~~~~~~~
A^{RR}_{d} = \frac{2}{3}\, t_W\,N_{11} \,,
\end{eqnarray}
for the down--type quarks $d$, $s$ and $b$. The chirality--preserving
coefficients $A^{LL , RR}_q$ originate from the gauge interactions, but the
chirality--flipping coefficients $A^{LR , RL}_q$ from the Yukawa interactions,
respectively.\\

Finally, the interactions of the neutral Higgs bosons to the lightest 
neutralinos involve both the neutralino mixing and the Higgs boson mixing. 
For the sake of notation we introduce the symbol $G_k$ defined in terms of 
the induced phase $\xi$, the neutralino diagonalization matrix $N$, and 
the Higgs diagonalization matrix $O$:
\begin{eqnarray}
\label{gk}
G_k \equiv (N_{12}-t_W\, N_{11} )
          \left[\, i\, (N_{13}\, s_{\beta} 
	       +N_{14}\, c_{\beta}\, e^{i \xi})\, O_{1k} 
	       +N_{13}\, O_{2k} + N_{14}\, O_{3k}\, e^{i \xi} \right].   
\end{eqnarray} 
With this abbreviation, the interaction Lagrangian for the coupling
of the Higgs boson to the lightest neutralino pair is cast into a simple 
form:  
\begin{eqnarray} 
\label{hxx}
 {\cal L}_{H\chi\chi} = \frac{e}{2 s_W} \sum_{k=1,2,3}
  \bar{\tilde{\chi}}^0_1\left[\,\real(G_k)+i\,\imag(G_k)\gamma_5\,\right]
       \tilde{\chi}^0_1\, H_k \,,
\end{eqnarray}
where $\real(G_k)$ and $\imag(G_k)$ denote the real and
imaginary parts of the coefficient $G_k$ in Eq.~(\ref{gk}), respectively.\\

It is now straightforward to derive the coefficients $A_q$ and $B_q$ involving
the effective four--fermion Lagrangian in the non--relativistic limit 
for the neutralino--quark elastic scattering:
\begin{eqnarray}
{\cal L}_{\rm eff}=A_q\,(\bar{\chi}_1\gamma^\mu\gamma_5 \chi_1)
                   \, (\bar{q} \gamma_\mu \gamma_5 q)
                +B_q\,(\bar{\chi}_1\chi_1)\, (\bar{q}q).
\end{eqnarray}
The $t$--channel $Z$--exchange diagram contributes to the spin--dependent 
part, and the $t$--channel Higgs--exchange diagrams to 
the spin--independent part, while the $s$--channel squark--exchange diagrams 
contribute to both the spin--dependent and spin--independent parts.
After appropriate Fierz transformations, the coefficients $A_q$ and $B_q$
in the non--relativistic limit read:  
\begin{eqnarray}
\label{eq:coefficients}
&& A_q = \frac{g^2}{16}
      \sum_{i=1,2} 
       \frac{\left|\, B_q^{i L}\right|^2 + \left|\, B_q^{iR}\right|^2 }
            { m_{\tilde{q_i}}^2 - (m_{\tilde{\chi_0}} -m_q )^2 }
  - \frac{G_F}{\sqrt{2}}
      \left[\left|N_{13}\right|^2 -\left|N_{14}\right|^2\right]
           T_3 ,\nonumber\\[2mm]
&& B_q = -\frac{g^2}{8}
       \sum_{i=1,2}
       \frac{\real (B_q^{iL} {B_q^{iR}}^*) }
            { m_{\tilde{q_i}}^2 - (m_{\tilde{\chi_0}}-m_q )^2 }
- \frac{g\, h_q }{2\sqrt{2}} 
       \sum_{k=1}^{3} \frac{\real(G_k)}{m_{H_k}^2} 
          \left\{\begin{array}{ll}
           O_{2k} &\textrm{for $q=d$}\\
           O_{3k} &\textrm{for $q=u$}
                  \end{array} \right . \,,
\end{eqnarray} 
where $G_F$ is the Fermi constant, and $h_q$ and $T_3$ are the Yukawa 
coupling and the third isospin component of the quark $q$, respectively. 
We have confirmed that in the CP--invariant theories, the expressions 
(\ref{eq:coefficients}) are consistent with those in \cite{Jungman:1996df}. 
Note that the coefficient $A_q$ contains the $Z$--exchange contribution
as well as the squark--exchange contributions, while the coefficient $B_q$ 
has the Higgs--exchange contributions as well as the squark--exchange 
contributions. The spin--dependent part is not suppressed by quark
masses so that it can be large for the lightest neutralino of higgsino type,
i.e, for large values $|N_{13}|$ and $|N_{14}|$. On the other hand, 
the spin--independent part from the $B_q$ terms is always 
proportional to quark masses and in particular the Higgs--exchange
contributions are sizable only when the lightest neutralino is
a well--balanced mixture of gaugino and higgsino states as seen from
Eq.~(\ref{gk}) and the size of the scalar--pseudoscalar mixing. 
Therefore, it might be naively expected that the small 
first and second generation quark masses will give rise to very small 
spin--independent cross sections. However, the spin--independent cross 
section can be significantly enhanced by coherent neutralino--quark 
scattering effects so that they dominate over the spin--dependent cross 
section for heavy nuclei.\\

\subsection{Neutralino--nucleus elastic scattering cross sections }

In order to obtain the scattering cross section of the lightest neutralino 
off heavy nuclei we need to know the detailed information on
the configuration of the protons and neutrons inside each heavy nucleus 
and on that of quarks and gluons inside each proton and neutron. 
In the present work, we will not touch on this issue in detail, but 
simply take $^{19}$F, $^{29}$Si, and $^{73}$Ge as three heavy nuclei 
and use the parametrizations of the form factors for 
each neutralino--nucleus scattering cross 
section as presented in Ref.~\cite{Jungman:1996df}.\\

The total cross section for each elastic scattering
process $\tilde{\chi}^0_1\, N \rightarrow \tilde{\chi}^0_1\, N$ with 
$N$$=$F, Si, and Ge at zero momentum transfers can be divided into
the axial--vector and scalar parts:
\begin{eqnarray}
\sigma = \sigma_{\rm A} + \sigma_{\rm S}.
\end{eqnarray}

First of all, let us consider the spin--dependent part that is parametrized
as follows:
\begin{eqnarray}
\sigma_{\rm A}=\frac{32}{\pi}\, G^2_F\, J(J+1)\,\, m^2_r\, \Lambda^2,
\end{eqnarray}
where $m_r$ is the neutralino--nucleus reduced mass 
$m_r=m_{\tilde{\chi}^0}m_N/(m_{\tilde{\chi}^0}+m_N)$ and $J$ 
the total angular momentum of the nucleus, the value of which is 
$1/2$ for ${^{19}}{\rm F}$ and ${^{29}}{\rm Si}$, $9/2$
for ${^{73}}{\rm Ge}$, respectively, and the spin-- and target--dependent 
quantity $\Lambda$ is given by
\begin{eqnarray}
\Lambda =\frac{1}{J}\left[\, a_p \, \langle S_p \rangle 
                            +a_n \, \langle S_n \rangle\right],
\end{eqnarray}
where $\langle S_p \rangle= \langle N|S_p |N \rangle $ 
and  $\langle S_p \rangle= \langle N|S_n |N \rangle $ and
are the expectation values of the spin content of the proton and neutron
group in the nucleus, respectively.\\

These expectation values $\langle S_{p,n}\rangle$ are different for 
each target nucleus and their explicit values for the target nuclei 
${^{19}}{\rm F}$, \, ${^{29}}{\rm Si}$ and ${^{73}}{\rm Ge}$ are given 
in the shell model by
\begin{eqnarray}
\label{eq:quantity 1}
&& \langle S_{p}\rangle_{_{\rm F}}\,\, =+0.415,\qquad\ \
   \langle S_{n}\rangle_{_{\rm F}}\,\, =-0.047, \nonumber \\
&& \langle S_{p}\rangle_{_{\rm Si}}\,=-0.002,\qquad \ \
   \langle S_{n}\rangle_{_{\rm Si}}\,=+0.130, \nonumber \\
&& \langle S_{p}\rangle_{_{\rm Ge}}=+0.011,\qquad \ \
   \langle S_{n}\rangle_{_{\rm Ge}}=+0.491. 
\end{eqnarray}
On the other hand, the coefficients $a_p$ and $a_n$ are parametrized 
in terms of the quark spin contents of the proton and neutron, 
$\Delta_q^p$ and $\Delta_q^n$, respectively, and the 
effective axial--vector couplings $A_q$ as 
\begin{eqnarray}
a_p = \sum_{q=u,d,s}{\frac{A_q}{\sqrt{2}\,G_{F}}}\,\Delta_q^p, \qquad
a_n = \sum_{q=u,d,s}{\frac{A_q}{\sqrt{2}\,G_{F}}}\,\Delta_q^n.
\end{eqnarray}
In the present work, we take the factors $\Delta_q^{p,n}$  $(q=u,\, d,\,
s)$ \cite{10} to be :    
\begin{eqnarray}
\label{eq:quantity 2}
&& \Delta_{u}^{p}=+0.77,\qquad 
   \Delta_{d}^{p}=-0.38,\qquad 
   \Delta_{s}^{p}=-0.09,\nonumber \\
&& \Delta_{u}^{n}=-0.38,\qquad 
   \Delta_{d}^{n}=+0.77,\qquad 
   \Delta_{s}^{n}=-0.09.
\end{eqnarray}
More detailed information on the expectation $\langle S_{p,n}\rangle$ 
in Eq.~(\ref{eq:quantity 1}) and the factors $\Delta_q^p$ and $\Delta_q^n$ 
in Eq.~(\ref{eq:quantity 2}) can be found in Ref.~\cite{Jungman:1996df}.\\

Secondly, the spin--independent cross section to which each
neutralino--proton or neutralino--neutron scattering contributes in a
coherent manner is parameterized as
\begin{eqnarray}
\sigma_{_{\rm S}}=\frac{4\,m_r^2}{\pi}\,\left[\,Zf_p+(A-Z)\,f_n\,\right]^2 , 
\end{eqnarray}
where $Z$ and $A$ denote the atomic number and mass number of the nucleus,
respectively. In the limit of $m_{\chi} \ll m_{\tilde{q}}$ and 
$m_{q} \ll m_{\tilde{q}}$, the effective couplings of the lightest
neutralino to protons and neutrons, $f_p$ and $f_n$, are approximated to be 
\begin{eqnarray}
\frac{f_{p,n}}{m_{p,n}}=\sum_{q=u,d,s}f_{_{{\rm T}_q}}^{p,n}\, \frac{B_q}{m_q}
                       +\frac{2}{27}\,f_{_{{\rm T}_G}}^{p,n}\sum_{q=c,b,t} 
		         \frac{B_q}{m_q},
\end{eqnarray}
up to the lowest order in  $1/m_{\tilde{q}}$, where the 
the parameters $f_{{\rm T}_q}^{p,n}$ are defined as
\begin{eqnarray}
f_{_{{\rm T}_q}}^{p,n}=\frac{\langle n,p|\,m_q \bar{q}q|n,p\rangle}{m_{n,p}},
                       \qquad
f_{_{{\rm T}_G}}^{p,n}=1-\sum_{q=u,d,s}f_{{\rm T}_q}^{p,n}.
\end{eqnarray}
In our present numerical analysis these quantities are taken to 
be   
\begin{eqnarray}
&& f_{_{{\rm T}_u}}^p=0.019,\qquad
   f_{_{{\rm T}_d}}^p=0.041,\qquad
   f_{_{{\rm T}_s}}^p=0.140,\nonumber \\
&& f_{_{{\rm T}_u}}^n=0.023,\qquad
   f_{_{{\rm T}_d}}^n=0.034,\qquad
   f_{_{{\rm T}_s}}^n=0.140.
\end{eqnarray}
as in Ref.~\cite{11}.\\

We note in passing that some contributions from the other interactions such
as so--called twist--2 operators are numerically small.

\section{Numerical results}
\label{sec:section 4}

We are now ready to present some numerical results.
It is known that loop--induced CP violation in the Higgs sector
can only be large if both $|\mu|$ and $|A_t|$ ( or $|A_b|$ if
$\tan\beta\gg 1$) are sizable. In addition, for moderate values of
$\tan\beta$ the contributions from the (s)bottom sector are still
quite small so that our numerical results are not sensitive to
$m_{\tilde{D}}$ and $A_b$. In light of these points, we
therefore choose
\begin{eqnarray}
\label{eq:parameter 1}
|A_t|=|A_b|= 2\, m_{\tilde{Q}} = 2\, m_{\tilde{U}}
           = 2\, m_{\tilde{D}}=1.0 \ \ {\rm TeV},
\end{eqnarray}
and take the phases of $A_t$ and $A_b$ to be equal. Numerically we find
that for the large squark masses in Eq.~(\ref{eq:parameter 1}) 
the squark--exchange contributions to the neutraluno--nucleus scattering
are very small compared to the other $Z$--boson and Higgs--exchange 
contributions.\\

Since the main focus will be on the effects of the three CP--violating 
phases $\{\Phi_1,\, \Phi_\mu,\, \Phi_A\}$ on the neutralino--nucleus 
scattering cross section, we do not perform any parameter scan
on all the real SUSY parameters in the present work. Nevertheless, 
it is crucial to cover the overall parameter space effectively.
So, except for the gaugino mass unification condition for the
moduli of the SU(2) and U(1) gaugino masses \footnote{The phase $\Phi_1$ 
of $M_1$ is zero up to the one--loop level for a positive $M_1$ 
in the minimal supergravity model.}
\begin{eqnarray}
|M_1| = \frac{5}{3}\, \tan^2 \theta_{\rm W}\, M_2 \,
        \approx\, 0.5\,\, M_2\, , 
\end{eqnarray}
we consider three different sets for the gaugino and higgsino mass
parameters:
\begin{eqnarray}
\label{eq:parameter set}
\begin{array}{lcl}
{\rm Gaugino} &:&M_2=100\ \ {\rm GeV},\quad |\mu|=1.0\ \ {\rm TeV},\\[2mm]
{\rm Higgsino}&:&M_2=500\ \ {\rm GeV},\quad |\mu|=100\ \ {\rm GeV},\\[2mm]
{\rm Mixed}   &:&M_2=|\mu|=150 \ \ {\rm GeV}.
\end{array}
\end{eqnarray}
As will be shown later, the lightest neutralino is Bino--like in the
gaugino case, while it is higgsino--like in the Higgsino case.\\

Considering the Higgs search experiments and the constraints on the
CP--phases of the top and bottom squark sectors from the electron and 
neutron EDMs at the two--loop level, we take two typical values of 
$\tan\beta$; $\tan\beta=3$ and 10. In addition, we take two values, 
$150$ and $500$ GeV for the pseudoscalar mass parameter $m_A$; 
In the former case, all the neutral Higgs bosons have
masses less than 200 GeV, while in the latter case two neutral Higgs bosons
are much heavier than the lightest Higgs boson. As a result, a significant
CP--violating Higgs--boson mixing is expected only for $m_A=150$ GeV. 
Finally, we take the running masses $\bar{m_t}(m_t)=165 ~{\rm GeV}$ and 
$\bar{m_b}(m_b)=4.2$ GeV as the top and bottom quark masses. We note in
passing that the spin--independent cross section is rather strongly 
dependent on these quark masses \cite{12}.\\

The loop--induced phase $\xi$ between the vacuum expectation values of 
two neutral Higgs bosons can be adjusted to be zero by taking an appropriate
renormalization scheme. With this adjustment and the universality assumption 
for the trilinear parameters, we can find that the Higgs--boson sector as well
as the top and bottom squark sectors involves only the combination 
$\Phi_{A\mu}\equiv\Phi_A+\Phi_\mu$ of the two rephasing--invariant phases
$\Phi_A$ and $\Phi_\mu$, while the light neutralino sector involves
both $\Phi_1$ and $\Phi_\mu$, but not $\Phi_A$.

\subsection{Mass spectra}

Based on the above parameter sets as in Eq.~(\ref{eq:parameter set}), 
let us first investigate the effects 
of the rephasing--invariant CP phases on the mass spectra for the
the lightest neutralino, top squarks and Higgs bosons.
Figure~2 shows the dependence of the lightest neutralino mass on the
phase $\Phi_\mu$ for five different values $\{0^0,\, 45^0,\,
90^0,\, 135^0,\, 180^0\}$ for the phase $\Phi_1$; the upper left (right)
frame is for $\tan\beta=3$ ($10$) in the gaugino case, the middle
left (right) frame is for $\tan\beta=3$ ($10$) in the higgsino case,
and the lower left (right) frame for $\tan\beta=3$ ($10$) in the 
mixed case. The approximate form of the lightest neutralino mass 
is given by 
\begin{eqnarray}
\label{eq:neutralino mass}
m_{\chi_0}\, \simeq\, \left\{\begin{array}{lll}
 {\rm Gaugino} &:&  \left|M_1\right|-\frac{m_z^2 s_W^2}{\left|\mu\right|}\,
                     \sin 2\beta \cos(\Phi_1+\Phi_\mu),\\[2mm]
 {\rm Higgsino} &:& \left|\mu\right|-\frac{m^2_Z}{2}\,(1+\sin 2\beta)\,
                    \left[\frac{s_W^2}{|M_1|}\cos(\Phi_1+\Phi_\mu)
                         +\frac{c_W^2}{M_2}\cos\Phi_\mu\right],
                         \end{array}\right.
\end{eqnarray}             
in the gaugino and higgsino cases, respectively. \\

\addtocounter{figure}{1}
\begin{center}
\begin{figure}[htb]
 \vspace*{0.1cm}
 \hspace*{2.0cm}
 \epsfxsize=12.0cm \epsfbox{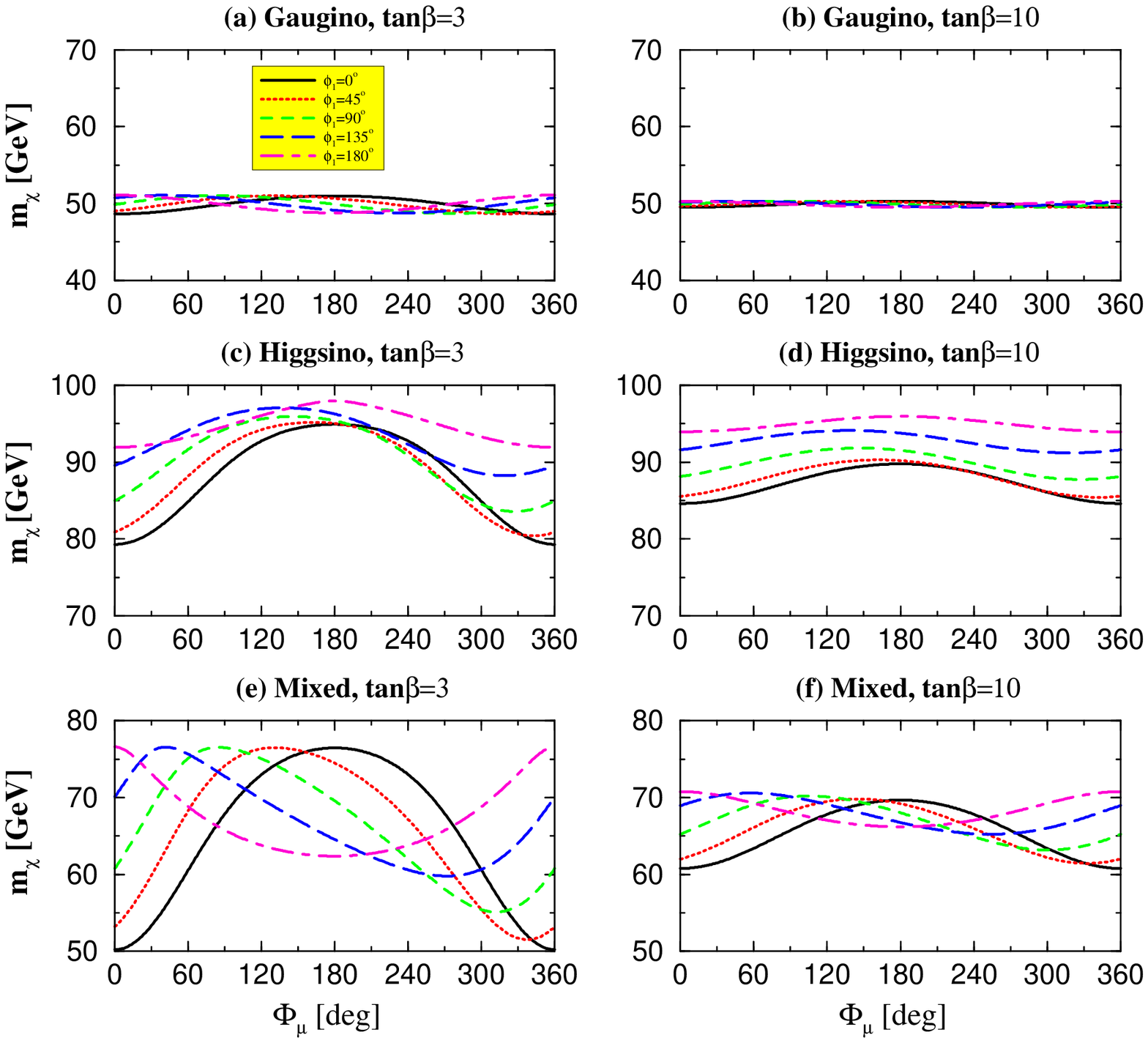}
 \vspace*{0.0cm}
 \caption{\it The dependence of the lightest neutralino mass 
              $m_{\tilde{\chi}^0_1}$ on the phase $\Phi_\mu$ 
	      for five different values $\{0^0,\, 45^0,\, 90^0,\, 
	      135^0,\, 180^0\}$ for the phase $\Phi_1$; 
	      the upper left (right) frame is for $\tan\beta=3$ ($10$) 
	      in the gaugino case, the middle left (right) frame is 
	      for $\tan\beta=3$ ($10$) in the higgsino case,
              and the lower left (right) frame for $\tan\beta=3$ ($10$) 
	      in the mixed case.}
 \label{fig:fig2}
 \vspace*{-0.5cm}
\end{figure}
\end{center}

Combined with the
approximate expressions (\ref{eq:neutralino mass}), Figure~\ref{fig:fig2}
shows several interesting features about the sensitivity of the
the neutralino mass to the phases $\Phi_A$ and $\Phi_\mu$:
\begin{itemize}
\item The mass is mainly determined by the modulus of $M_1$ and $\mu$ in 
      the gaugino and higgsino cases, respectively, while in the mixed case
      the mass is affected by both $M_1$ and $\mu$.
\item As the phase--dependent parts are accompanied by a factor $\sin 2\beta$
      as shown in the expressions, the sensitivity of the neutralino mass
      to the phases is smaller for larger $\tan\beta$ in every case.
\item In the mixed case, the sensitivity of the neutralino mass to the 
      phase $\Phi_1$ is larger when $\Phi_\mu=0$ or $2\,\pi$.
\end{itemize}
Consequently, it is expected that only in the mixed and higgsino cases with
relatively small $|M_1|$ and $|\mu|$ the neutralino--nucleus scattering
cross sections can be affected by the phases $\Phi_\mu$ and
$\Phi_1$ through the lightest neutralino mass $m_{\tilde{\chi}^0_1}$.\\

\begin{center}
\begin{figure}[htb]
 \vspace*{0.0cm}
 \hspace*{2.0cm}
 \epsfxsize=12.0cm \epsfbox{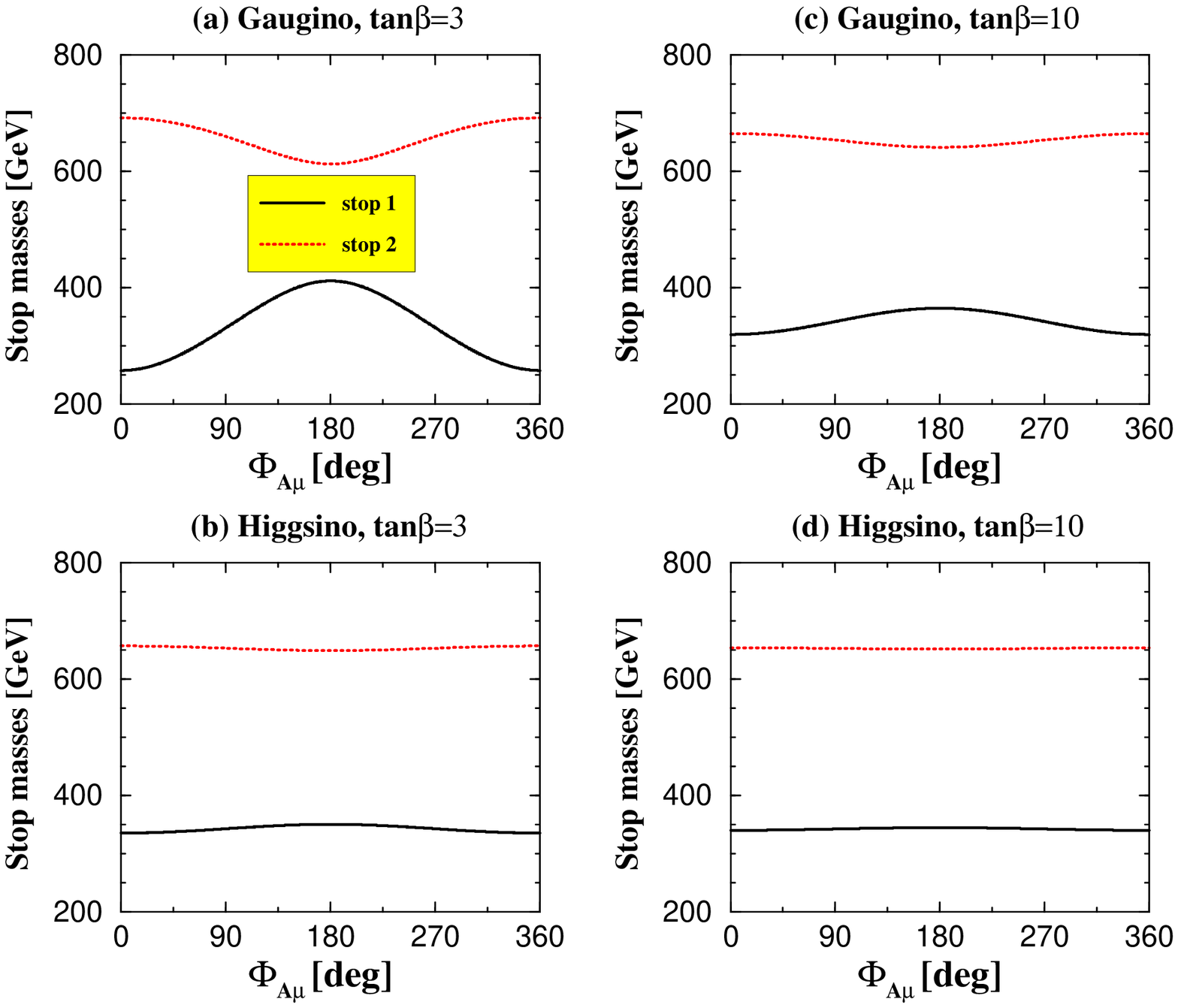}
 \vspace*{0.0cm}
 \caption{\it The dependence of the light and heavy top squark masses
              $m_{\tilde{t}_{1,2}}$ on the phase $\Phi_{A\mu}$ 
	      for (a) $|\mu|=1$ TeV and $\tan\beta=3$, (b)
	      $|\mu|=100$ GeV and $\tan\beta=3$, (c) $|\mu|=1$ TeV and 
	      $\tan\beta=10$, (d) $|\mu|=100$ GeV and $\tan\beta=10$
	      with the parameter set (\ref{eq:parameter 1}) for the 
	      squark mass parameters and trilinear parameters.}
 \label{fig:fig3}
 \vspace*{-0.5cm}
\end{figure}
\end{center}

For moderate values of $\tan\beta$ the loop--induced effects to the neutral
Higgs boson sector is dominated by the top (s)squark contributions 
due to the largest top Yukawa coupling. In this light, we present 
in Fig.~\ref{fig:fig3} the dependence of the top squark masses 
$m_{\tilde{t}_{1,2}}$ on the 
phase $\Phi_{A\mu}$ for (a) $|\mu|=1$ TeV and $\tan\beta=3$, 
(b) $|\mu|=100$ GeV and $\tan\beta=3$, (c) $|\mu|=1$ TeV and $\tan\beta=10$, 
(d) $|\mu|=100$ GeV and $\tan\beta=10$ with the parameter set 
(\ref{eq:parameter 1}) for the squark mass parameters 
and trilinear parameters. Analytically, we find that the phase dependence
of the top squark masses is determined by the quantity:
\begin{eqnarray}
\label{eq:stop mass splitting}
\Delta_{12}=m_t\, \imag(A_t\,\mu\,{\rm e}^{i\xi})\, \cot\beta.
\end{eqnarray}
Figure~\ref{fig:fig3} and Eq.~(\ref{eq:stop mass splitting}) clearly show
that the top squark masses are strongly dependent on the phase 
$\Phi_{A\mu}$ only in the gaugino case with large $|\mu|$ and small 
$\tan\beta$. In the other cases, the top squark masses are almost
insensitive to the phase $\Phi_{A\mu}$.

\begin{center}
\begin{figure}[htb]
 \vspace*{0.1cm}
 \hspace*{2.0cm}
 \epsfxsize=12.0cm \epsfbox{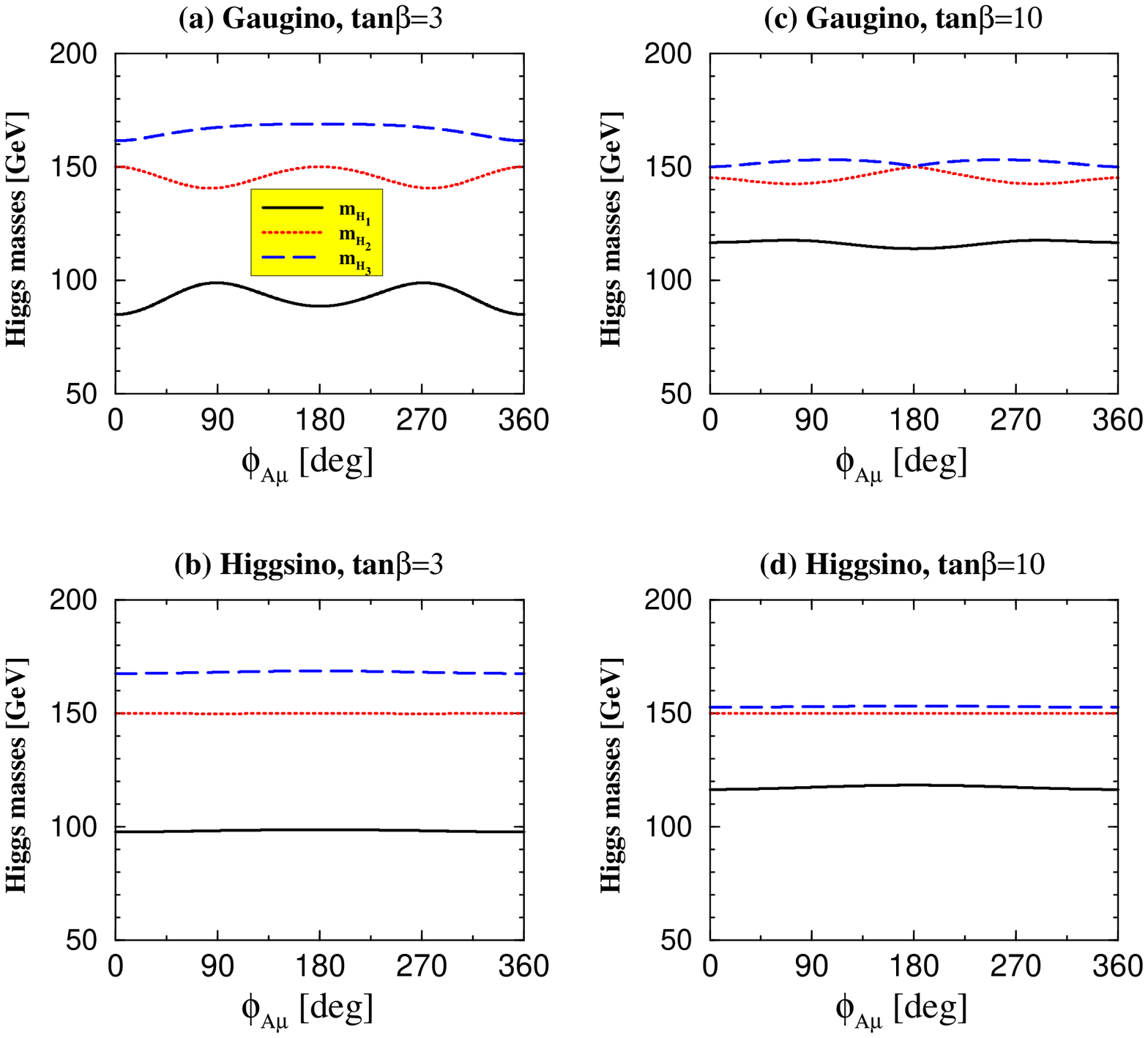}
 \vspace*{0.0cm}
 \caption{\it The Higgs--boson mass spectrum with respect to the
              phase $\Phi_{A\mu}$ in (a) the gaugino case and (b)
	      the higgsino case with $\tan\beta=3$, and  
              in (c) the gaugino case and (d) the higgsino case with 
	      $\tan\beta=10$; the pseudoscalar mass parameter $m_A$ is
	      set to be 150 GeV.}
 \label{fig:fig4}
 \vspace*{0.0cm}
\end{figure}
\end{center}

The mass spectrum of the neutral Higgs bosons is sensitive to the
combination $\Phi_{A\mu}=\Phi_A+\Phi_\mu$. So, we exhibit in Fig.~3 
the Higgs--boson mass spectrum with respect to the phase $\Phi_{A\mu}$ 
with $m_A=150$ GeV. The mass difference between the 
the lightest Higgs boson and the heavier Higgs boson is more significant
for smaller $\tan \beta$. The reason is that for the pseudoscalar mass
comparable to $m_Z$, the off--diagonal entries of the Higgs--boson 
mass matrix are essentially proportional to $\sin 2 \beta$.
On the other hand, the sensitivity of the Higgs--boson masses to the 
phase $\Phi_{A\mu}$ is strongly suppressed for large $m_A$ due to the
suppressed mixing. Because of this feature we have shown the 
Higgs mass spectrum only for $m_A=150$ GeV in Fig.~3.
For moderate values of $\tan\beta$ the size of the one--loop induced 
CP--violating neutral Higgs--boson mixing is dictated by the factor
\begin{eqnarray}
\Delta_{\tilde{t}} = \frac{\imag (A_t \mu \, {\rm e}^{i\xi})}{
                           m^2_{\tilde{t}_2}-m^2_{\tilde{t}_1}}.
\end{eqnarray}
Therefore, only when both $|A|$ and $|\mu|$ are large, the Higgs boson masses
become sensitive to the phase $\Phi_{A\mu}$. This feature is clearly
reflected in Fig.~3 for the Higgs mass spectrum; in the gaugino case 
with a large value of $|\mu|=1$ TeV, the Higgs boson masses depend strongly on
the phase, while in the higgsino case with a small value of $|\mu|=100$ GeV,
the Higgs boson masses are almost insensitive to the phase.
Consequently, the large effects of the phase $\Phi_{A\mu}$ on the Higgs
masses and mixing are expected for large $|\mu|$ as well as small $m_A$ and
$\tan\beta$.

\subsection{Neutralino--nucleus scattering cross sections}

With the previous comprehensive investigations of the mass spectra 
for the lightest neutralino, top squarks and neutral Higgs bosons, 
let us study in this section the dependence of the neutralino--nucleus 
cross sections on the CP--violating phases $\{\Phi_1,\Phi_\mu,\Phi_A\}$ 
as well as the other real SUSY parameters.\\

Firstly, we evaluate the elastic cross sections in the CP--invariant theories.
We take the three phases to be vanishing; $\Phi_A=\Phi_{\mu}=\Phi_1=0$
and consider F, Si and Ge as three typical target nuclei.
Table~1 shows the neutralino--nucleus elastic scattering cross sections;
spin--independent and spin--dependent parts, $\sigma_S$ and
$\sigma_A$,\, for the three different scenarios and two values of 
$m_A$; the upper table is for $\tan\beta=3$ and the lower part for 
$\tan\beta=10$. Note that the size of the cross sections depends
significantly on the scenarios and on the target nuclei as well as on
the pseudoscalar mass parameter $m_A$:
\begin{itemize}
\item The cross sections are strongly suppressed in the gaugino case, 
      but they are very much enhanced in the mixed and higgsino cases.
      This suppression feature is more predominant in the spin--dependent
      cross section, for the coupling of the $Z$ boson to the neutralinos 
      is strongly suppressed in the gaugino case, i.e. the
      absolute magnitudes of $N_{13}$ and $N_{14}$ are extremely small.
\item The spin--independent cross sections are much more enhanced for 
      heavier nucleus and become dominant over the spin--dependent cross 
      section as can be expected from the coherent scattering effects.
\item The spin--independent cross sections are strongly suppressed for
      large pseudoscalar mass $m_A$ and large $\tan\beta$, especially 
      in the gaugino case.
\end{itemize}
The elastic scattering process $\tilde{\chi}^0_1\, q
\rightarrow \tilde{\chi}^0_1\, q$ is related to the neutralino--neutralino
annihilation process, $\tilde{\chi}^0_1\tilde{\chi}^0_1\rightarrow
q\bar{q}$, contributing to the relic density of the SUSY dark matter.
We note that in most cases the gaugino scenario is favored by the
estimates of the relic density.\\

\begin{table}[h]
\caption{\label{tab:table1}
 {\it The neutralino--nucleus elastic scattering cross sections;
       spin--independent and spin--dependent parts for the
       three different scenarios and two values of $m_A$ in the
       CP--invariant case, $\Phi_\mu=\Phi_1=\Phi_A=0$. {\rm F,\, Si}
       \, and {\rm Ge} are considered as three different target nuclei; 
       the upper table is for $\tan\beta=3$ and the lower part for 
       $\tan\beta=10$.}}
       \mbox{ }\\[-3mm]
\begin{center}
\begin{tabular}{|l c|c c|cc|cc|}\hline 
 \multicolumn{2}{|c|}{$\tan \beta=3$}
&\multicolumn{2}{c|} {$\rm F$}
&\multicolumn{2}{c|} {$\rm Si$}
&\multicolumn{2}{c|} {$\rm Ge$}\\ 
& $m_A$ [\,GeV]           &$\sigma_{_{\rm S}}$[fb] &$\sigma_{_{\rm A}}$[fb]
& $\sigma_{_{\rm S}}$[fb] &$\sigma_{_{\rm A}}$[fb]
& $\sigma_{_{\rm S}}$[fb] &$\sigma_{_{\rm A}}$[fb]\\[1mm] \hline\hline
Gaugino  &$150$ &0.17 &0.02 &0.63 &0.002 &11.6 &0.028  \\   
         &$500$ &0.15 &0.02 &0.57 &0.002 &10.6 &0.028 \\ \hline
Higgsino &$150$ &10.7 &77.0 &43.3 &8.79  &992  &161 \\   
         &$500$ &11.3 &77.0 &45.8 &8.79  &1049 &161 \\ \hline
Mixed    &$150$ &25.0 &120  &94.7 &12.8  &1769 &191 \\   
         &$500$ &26.2 &120  &99.1 &12.8  &1852 &191 \\ \hline
\end{tabular}

\vskip 0.6cm
\begin{tabular}{|l c|c c|cc|cc|}\hline
 \multicolumn{2}{|c|}{$\tan \beta=10$}
&\multicolumn{2}{c|} {$\rm F$}
&\multicolumn{2}{c|} {$\rm Si$}
&\multicolumn{2}{c|} {$\rm Ge$}\\ 
& $m_A$ [\,GeV]        &$\sigma_{\rm S}$[fb] &$\sigma_{\rm A}$[fb]
& $\sigma_{\rm S}$[fb] &$\sigma_{\rm A}$[fb]
& $\sigma_{\rm S}$[fb] &$\sigma_{\rm A}$[fb]\\ \hline\hline
Gaugino  &$150$ &0.38  &0.02 &1.45 &0.002 &26.9  &0.035  \\   
         &$500$ &0.003 &0.02 &0.01 &0.002 &0.20  &0.035 \\ \hline
Higgsino &$150$ &59.7  &114  &244  &13.1  &5741  &247 \\   
         &$500$ &60.3  &114  &246  &13.1  &5800  &247 \\ \hline
Mixed    &$150$ &160   &157  &622  &17.3  &12689 &282 \\   
         &$500$ &161   &157  &626  &17.3  &12783 &282 \\ \hline
\end{tabular}

\end{center}
\end{table}
\begin{center}
\begin{figure}[htb]
 \vspace*{0.1cm}
 \hspace*{2.0cm}
 \epsfxsize=12.0cm \epsfbox{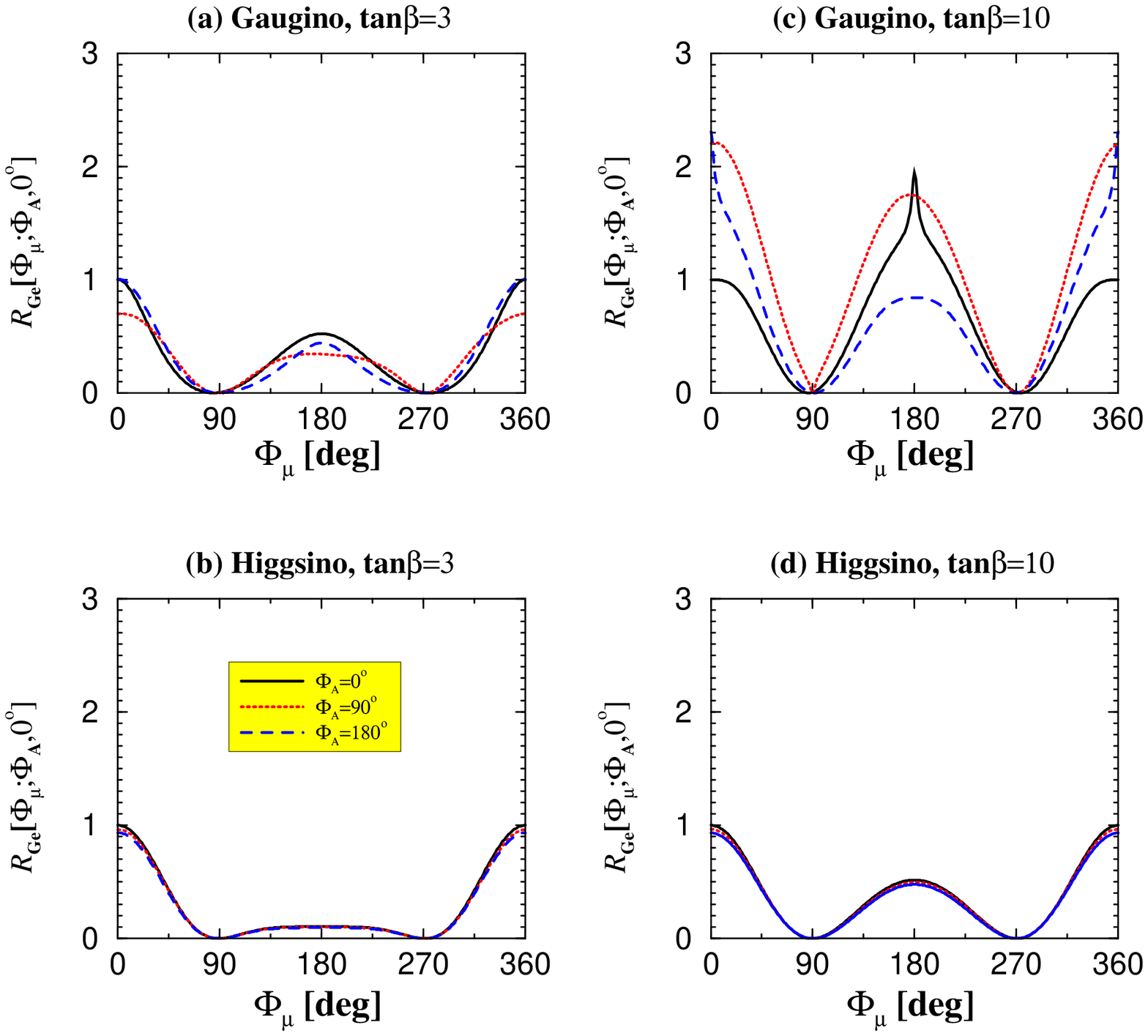}
 \vspace*{0.0cm}
 \caption{\it The ratio ${\cal R}_{_{\rm Ge}}[\Phi_\mu;\Phi_A,\Phi_1]$
	      with respect to the phase $\Phi_\mu$ for the parameter
	      set (\ref{eq:parameter 1}) and $m_A=150$ GeV
              in (a) the gaugino case and (b) the higgsino case 
	      with $\tan\beta=3$, and in (c) the gaugino case and 
	      (d) the higgsino case with $\tan\beta=10$.
	      Here, the phase $\Phi_1$ is taken to be $0^0$.}
 \label{fig:fig5}
 \vspace*{0.0cm}
\end{figure}
\end{center}

Secondly, we study the effects of the CP--violating phases on the
neutralino--nucleus scattering cross sections. Note that
the spin--dependent cross sections $\sigma_{_A}$ are independent of 
$\Phi_A$ and  almost independent of the phases $\Phi_1$ and
$\Phi_\mu$, especially in the gaugino and higgsino cases, because
the neutralino mixing elements $N_{13}$ and $N_{14}$  determining the
coupling of the $Z$ boson to the lightest neutralinos are
suppressed by the factors $m_Z/|\mu|$ and $m_Z /|M_{1,2}|$ in the 
gaugino and higgsino cases, respectively.
Even in the mixed case the lightest neutralino is almost Bino--like because of
the gaugino mass unification condition so that this scenario also
causes the spin--dependent cross sections are almost independent
of the CP--violating phases. Taking into account these points, we 
consider the ratio of the spin--independent cross sections only for 
the heaviest target nucleus Ge :
\begin{eqnarray}
\label{eq:ratio}
{\cal R}_{_{\rm Ge}}\,[\Phi_\mu;\Phi_A,\Phi_1]
   =\frac{\sigma_{_{\rm S}}[\, \Phi_\mu;\Phi_A,\Phi_1]}{
          \sigma_{_{\rm S}}[\,\,0\,\,;\,0\,,\, 0]}
\end{eqnarray}
for several values of the phases $\{\Phi_1,\, \Phi_A\}$ and
present the numerical results for the spin--independent
cross section ratio (\ref{eq:ratio}) by taking $m_A=150$ GeV
and the parameter set (\ref{eq:parameter 1}) in three figures; the first 
figure is for $\Phi_1=0^0$ (Figure~5), 
the second one for $\Phi_1=90^0$ (Figure~6) and the third one 
for $\Phi_a=180^0$ (Figure~7), respectively. Each figure contains four
frames for (a) the gaugino case and $\tan\beta=3$, (b) the higgsino case 
and $\tan\beta=3$, (c) the gaugino case and $\tan\beta=10$ and
(d) the higgsino case and $\tan\beta=10$; in each frame the solid line is
for $\Phi_A=0^0$, the dotted line for $\Phi_A=90^0$ and
the dashed line for $\Phi_A=180^0$.

\begin{center}
\begin{figure}[htb]
 \vspace*{0.5cm}
 \hspace*{2.0cm}
 \epsfxsize=12.0cm \epsfbox{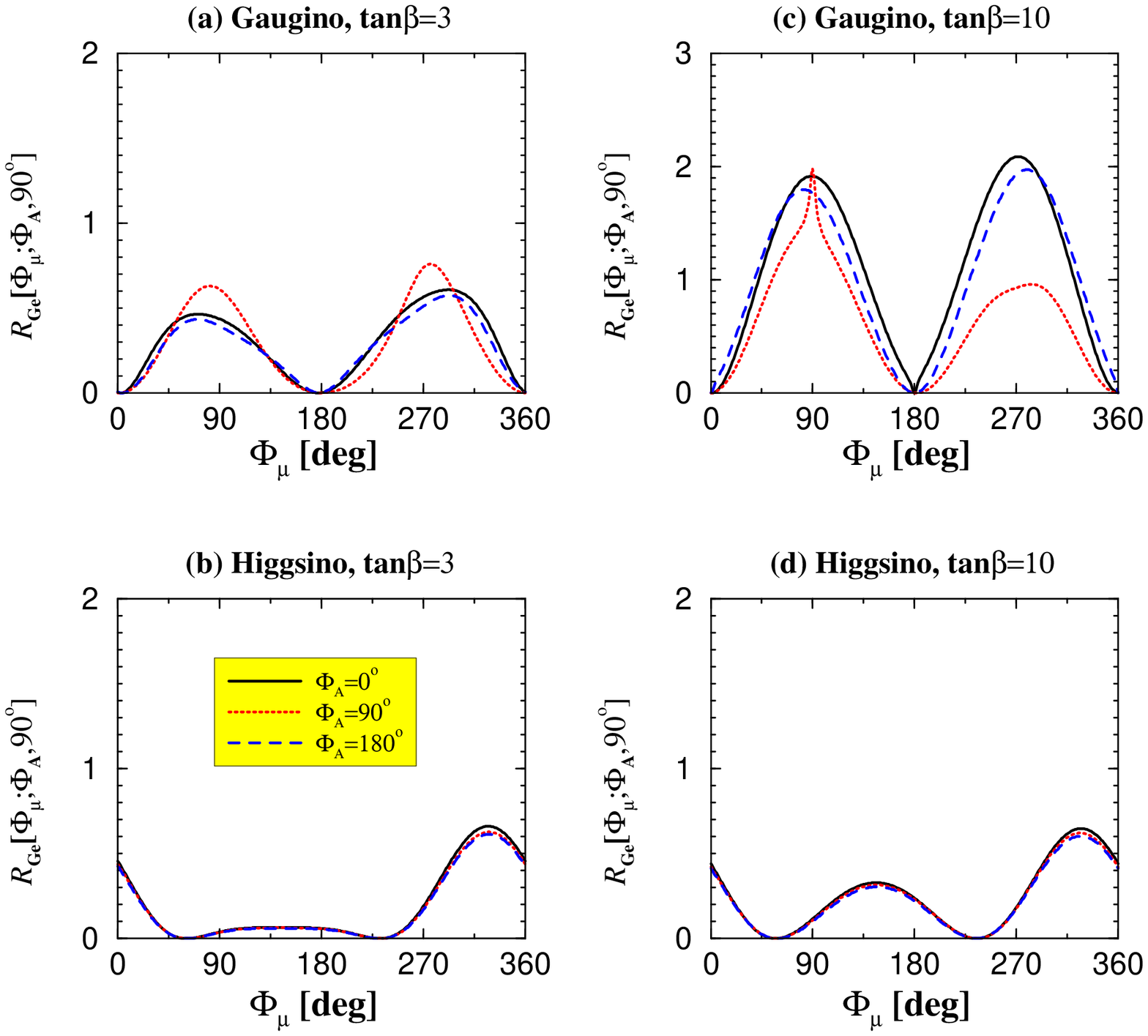}
 \vspace*{0.0cm}
 \caption{\it The same ratio ${\cal R}_{_{\rm Ge}}[\Phi_\mu;\Phi_A,90^o]$
              as in Fig.~\ref{fig:fig5} but for $\Phi_1=90^0$.}
 \label{fig:fig6}
 \vspace*{-0.7cm}
\end{figure}
\end{center}

\begin{center}
\begin{figure}[htb]
 \vspace*{0.5cm}
 \hspace*{2.0cm}
 \epsfxsize=12.0cm \epsfbox{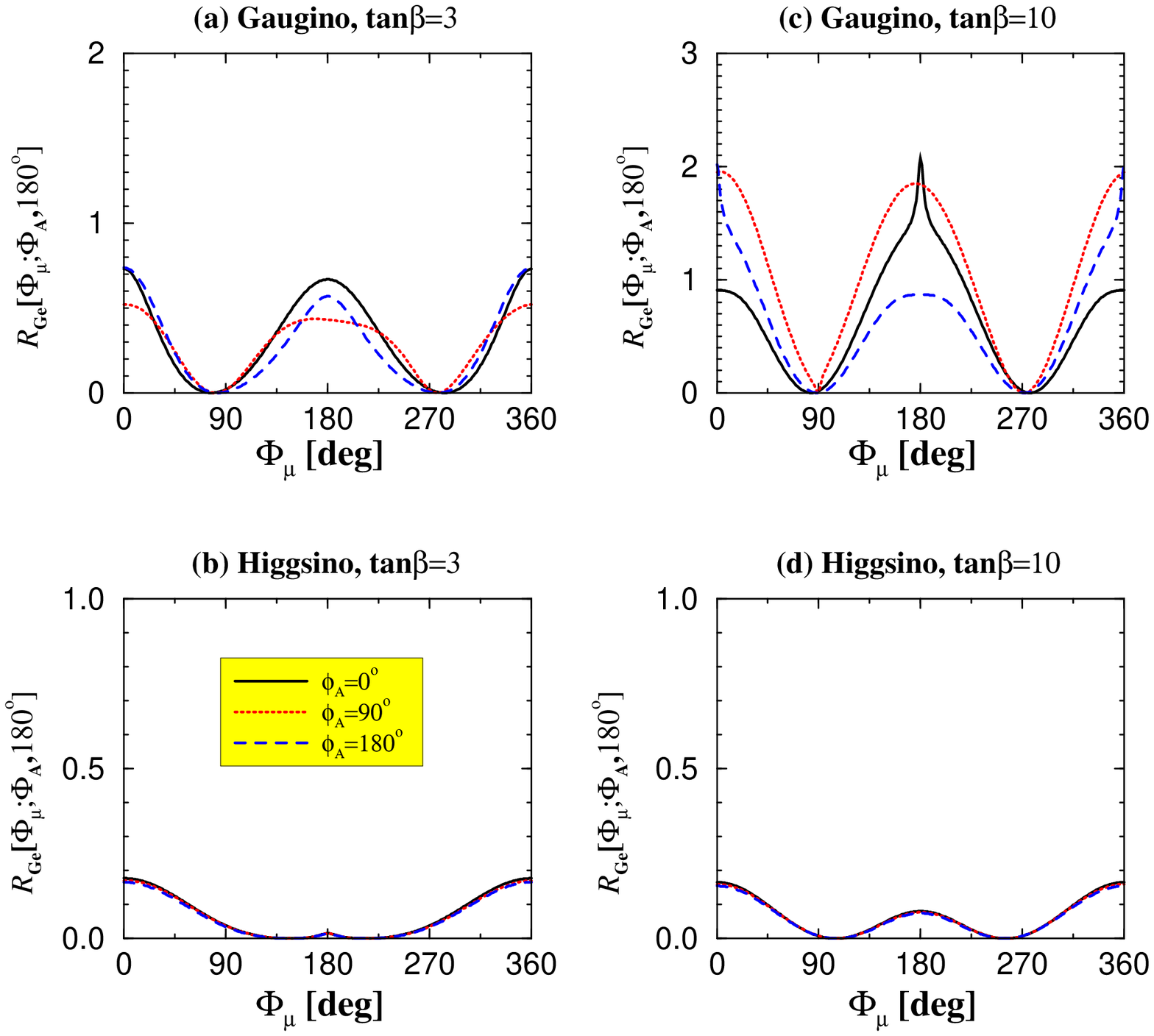}
 \vspace*{0.0cm}
 \caption{\it The same ratio ${\cal R}_{_{\rm Ge}}[\Phi_\mu;\Phi_A,\Phi_1]$
              as in Figs.~\ref{fig:fig5} and \ref{fig:fig6} but for 
	      $\Phi_1=180^0$.}
 \label{fig:fig7}
 \vspace*{1.0cm}
\end{figure}
\end{center}

Comparing the results presented in the three figures, we find several
interesting features about the dependence of the spin--independent ratio
${\cal R}_{_{\rm Ge}}$:
\begin{itemize}
\item In all the figures and all the scenarios the spin--independent 
      ratio is strongly dependent on the phase $\Phi_\mu$; it is striking
      that certain values of $\Phi_\mu$ render the cross sections 
      almost vanishing. 
\item The ratio is very sensitive to the phase $\Phi_A$ in the
      gaugino case, where the Higgs--boson couplings to the lightest 
      neutralinos as well as the quarks are significantly
      affected by the phases $\Phi_A$ and $\Phi_\mu$ as well as the 
      phase $\Phi_1$. This $\Phi_A$ dependence is expected to be
      more significant for small $m_A$ due to a larger mixing 
      between the scalar and pseudoscalar Higgs bosons. 
      The ratio itself varies more significantly with the phases
      $\Phi_\mu$ and $\Phi_A$ for $\tan\beta=10$ than $\tan\beta=3$.
\item In the higgsino case, the cross section ratio ${\cal R}_{\rm Ge}$
      is (almost) always suppressed for non--trivial values of $\Phi_\mu$, 
      i.e. the neutralino--nucleus cross section is maximal near 
      $\Phi_\mu=0$ and $360^o$. 
\item On the contrary, the cross section ratio is enhanced or suppressed 
      depending on the values of the phases $\{\Phi_1,\,\Phi_\mu,\, 
      \Phi_A\}$ in the gaugino case. Interestingly, the cross section
      ratio is maxmal when the sum of the phases $\Phi_\mu$ and
      $\Phi_1$ is $0^o$ ($360^o$)  or $180^o$, while the ratio is almost
      vanishing when the sum of the phases $\Phi_\mu$ and
      $\Phi_1$ is $90^o$ or $270^o$.  This feature can be understood from the
      the fact that in the gaugino case the matrix elements 
      ${\cal N}_{1k}$ ($k=1$ - 4) as well as the lightest neutralino 
      mass depend on only the rephasing--invariant combination 
      $\Phi_\mu+\Phi_1$ with a good approximation.
\item Comparing the lines for three different values of $\Phi_A$ in each
      frame of Figs.~\ref{fig:fig5}, \ref{fig:fig6}, and 
      \ref{fig:fig7} in the gaugino case, we can confirm that the phase
      determining the CP--violating Higgs--boson mixing is indeed the 
      rephasing--invariant combination $\Phi_{A\mu}$ of the phases 
      $\Phi_A$ and $\Phi_\mu$.
\end{itemize}

All these features can be understood by noting that the spin--independent 
interactions are primarily determined by the coupling strengths of the
Higgs bosons to the neutralinos and the quarks given in 
Eqs.~(\ref{eq:higgs-quark}) and (\ref{hxx}). The scalar couplings of a 
Higgs bosons to quarks can vanish when the Higgs boson becomes 
(almost) a CP--odd state for certain values of $\Phi_{A\mu}$ through 
the CP--violating scalar--pseudoscalar mixing. Certainly, for some values of
the phases the scalar couplings can be enhanced compared to the CP--invariant
case. On the other hand, the scalar couplings of the Higgs bosons to the 
lightest neutralinos, which is given by the real part of
$G_k$ in Eq.~(\ref{gk}), involves not only the CP--violating Higgs--boson
mixing but also the neutralino mixing. Therefore, these couplings are
expected to be strongly dependent on all the phases in the gaugino case.\\

To summarize, the CP--violating phases could reduce or enhance the 
neutralino--nucleus cross sections compared to those in the CP--invariant
theories. The phase $\Phi_\mu$ as well as the phase $\Phi_1$ affects the 
cross section significantly in most cases, while 
the phase $\Phi_A$ modifies the (spin--independent) cross section 
in the gaugino case with large $|\mu|$ through the phase combination
$\Phi_{A\mu}=\Phi_A+\Phi_\mu$ for relatively small $m_A$. 

%

\section{Summary and conclusions}
\label{sec:section 5}

In this paper, we have made a comprehensive investigation of 
the effects of the CP--violating phases on the neutralino--nucleus 
elastic scattering cross sections in the framework of the MSSM through
the neutralino mixing and the one--loop induced Higgs--boson mixing
from mainly the top (s)quark sector due to the largest top Yukawa
coupling. For the sake of numerical analysis, we have imposed the 
universal relation for the squark mass parameters and the gaugino mass
unification condition only for the moduli of the SU(2) and U(1) 
gaugino masses. We have taken large sfermion masses and moderate values of 
$\tan\beta$, allowing the CP--violating phases to have large
non--trivial values without violating the 
stringent constraints of the electron and neutron EDM measurements, 
as well as large trilinear 
parameters for a significant neutral Higgs--boson mixing.\\

Based on three scenarios -- gaugino, higgsino and mixed -- for 
the neutralino sector, we have performed a detailed numerical analysis for
the mass spectra of the lightest neutralino, top squarks and 
neutral Higgs bosons. The lightest neutralino mass is strongly dependent
on the phases $\Phi_1$ and $\Phi_\mu$ for small $\tan\beta$ and in the mixed
case with comparable gaugino and higgsino mass parameters. On the contrary,
the top squark and Higgs boson masses are significantly affected by
the combination $\Phi_{A\mu}$ of the phases $\Phi_\mu$ and $\Phi_A$
in the gaugino case with large $|\mu|$ and with small $\tan\beta$.\\

After estimating the neutralino--nucleus scattering cross sections
for the target nuclei, F, Si and Ge for the three scenarios in 
Eq.~(\ref{eq:parameter set}) in the CP--invariant theories, we have
studied the effects of the CP--violating phases on
the spin--independent cross section ratio ${\cal R}_{_{\rm Ge}}$
for the target nucleus Ge, for which the spin--independent cross
section is significantly enhanced by large coherent scattering
effects due to its large mass number. We have found that the CP--violating 
phases could reduce or enhance the cross section ratio significantly. 
The phase $\Phi_\mu$ as well as the phase $\Phi_1$ affects the ratio 
section in almost all the cases, while the phase $\Phi_A$ modifies 
the (spin--independent) cross section in the gaugino case with large 
$|\mu|$ through the phase combination $\Phi_{A\mu}=\Phi_A+\Phi_\mu$,
when the pseudoscalar mass $m_A$ is relatively small. \\

Consequently, the CP--violating
phases as well as all the real parameters in the MSSM are very crucial 
in estimating the neutralino--nucleus cross sections and so in determining
the possibility of experimentally detecting the lightest neutralinos if
they indeed constitute a major component of cold dark matter in the 
Universe.

\section*{Acknowledgements}

The work of S.Y.C was supported by Korea Research Foundation Grant 
(KRF--2000--015--DS0009) and the work of S.C.P by the BK21 Program.
And the work of J.H.J was in part by grant No. 20005--111--02-2
from the Korea Science and Engineering Foundation.


\end{document}